\documentclass[11pt]{article}

\usepackage[margin=1in]{geometry}

\usepackage{siunitx}
\usepackage{upgreek}
\usepackage{graphicx}
\usepackage{amsmath}
\usepackage{amssymb}
\usepackage{xfrac}
\usepackage{xcolor}
\usepackage{soul}

\usepackage{bm}

\usepackage{mathptmx}

\usepackage[super,comma,sort&compress]{natbib}

\usepackage{titlesec}
\titleformat{\section}{\Large\bfseries}{\thesection.}{1em}{}[]
\titleformat{\subsection}{\normalsize\bfseries}{\thesubsection.}{1em}{}[]
\titleformat{\subsubsection}{\normalsize\itshape}{\thesubsubsection.}{1em}{}[]
\titlespacing{\section}{0pt}{5pt}{0pt}[0pt]
\titlespacing{\subsection}{0pt}{3pt}{0pt}[0pt] 
\titlespacing{\subsubsection}{0pt}{0pt}{0pt}[0pt] 

\usepackage[font=footnotesize,labelfont=bf]{caption} 

\usepackage{float}
\usepackage[color=blue!15, size=footnotesize]{todonotes}

\usepackage{amsmath}
\usepackage{stackengine}

\begin{document}

\begin{center}

\Large

\textbf{Measurement of tissue viscosity to relate force and motion in collective cell migration}

\vspace{11pt}

\normalsize
Molly McCord$^{1,2}$, Jacob Notbohm$^{1,2,*}$
\vspace{11pt}

$^1$Department of Mechanical Engineering, University of Wisconsin--Madison, Madison, WI, USA\\
$^2$Biophysics Program, University of Wisconsin--Madison, Madison, WI, USA

$^*$ Contact: jknotbohm@wisc.edu

\end{center}

\vspace{0.3in}



\section*{Abstract}

In tissue development, wound healing, and cancer invasion, coordinated cell motion arises from active forces produced by the cells. The relationship between force and motion remains unclear, however, because the forces result from a sum of contributions from activity and the constitutive response of the cell collective. Here, we develop a method to decouple the forces due to activity from those due to constitutive response. As a model of an epithelial tissue, we use a monolayer of epithelial cells in the fluid state, for which the constitutive behavior is that of a viscous fluid. By careful study of the distribution of the ratio between shear stress and strain rate, we show that the order of magnitude of viscosity within the epithelial tissue is 100 Pa$\cdot$hr and that increasing (decreasing) the actomyosin cytoskeleton and cell-cell adhesions increase (decrease) the magnitude of tissue viscosity. These results establish tissue viscosity as a meaningful way to describe the mechanical behavior of epithelial tissues, and demonstrate a direct relationship between tissue microstructure and material properties. By providing the first experimental measurement of tissue viscosity, our study is a step toward separating the active and constitutive components of stress, in turn clarifying the relationship between force and motion and providing a new means of identifying how active cell forces evolve in space and time.

\section{Introduction}

Collective cell migration is a fundamental biological process that drives tissue development, wound healing, and cancer invasion.\cite{friedl2009, li2015, park2016, scarpa2016} In contrast to single-cell migration, collective cell migration gives rise to large-scale, complex patterns of motion, including swirling vortices, oscillations, velocity-aligned flocking, and transitions between solid-like and fluid-like states.\cite{poujade2007, serra-picamal2012, park2015, notbohm2016, malinverno2017, mongera2018, laang2024, shen2025} These complex behaviors arise from an interplay wherein biological activity and mechanical responses are linked. Forces have direct physical effects, such as by producing migration, and also control biological behaviors such as proliferation, differentiation, and gene expression.\cite{vogel2006, ladoux2017} While the forces produced by the cells can be experimentally measured---using techniques such as traction force microscopy and monolayer stress microscopy\cite{dembo1998, butler2002, trepat2009, tambe2011, tambe2013}---their direct comparison to motion remains a challenge as the forces measured are a combination of active (produced by the cells) and viscoelastic contributions.\cite{alert2020review}

The connection between force and motion is clear in theoretical models, wherein the total force in the monolayer is typically decomposed into a combination of viscoelastic components, which reflect the rheological response of the monolayer, and active components, which arise from cellular force generation.\cite{garcia2015, zimmermann2016, bi2016, barton2017} For example, the traction applied by cells to the underlying substrate, $\bm{t}$ can be written as $\bm{t} = \bm{t}^f+\bm{t}^a$, where $\bm{t}^f$ and $\bm{t}^a$ are frictional and active components of traction, respectively. Stresses $\bm{\sigma}$ can be written as $\bm{\sigma} = \bm{\sigma}^{ve}+\bm{\sigma}^a$, with $\bm{\sigma}^{ve}$ and $\bm{\sigma}^a$ being the viscoelastic and active contributions. The non-active components are directly related to cell motion or deformation; for example, the frictional component of traction is often written as $\bm{t^f}= \xi \bm{v}$, where $\xi$ is a cell-to-substrate friction constant, and $\bm{v}$ is the velocity.
The form of the viscoelastic component of stress depends on the choice of whether to interpret the system as primarily fluid or solid. In this manuscript, we treat the cell monolayer as a fluid, which is justified based on the time scales for turnover of actin of 14 s and E-cadherin of 4 min and the relaxation time for epithelial tissues of $\sim 1$ min.\cite{khalilgharibi2019, debeco2009,harris2012} These time scales are small compared to the typical time for an epithelial cell to migrate its body length ($\sim$3 hr). Thus, we assume the viscoelastic component of stress to be primarily viscous, writing it as  $\sigma^{ve} = \eta \dot{\varepsilon}_s$, where $\dot{\varepsilon}_s$ is the shear strain rate and $\eta$ is the tissue viscosity. Although $\eta$ is not directly related to activity, it can be altered indirectly by cellular activity, such as by activity-driven changes to the cytoskeleton and cell-cell adhesions.\cite{oriola2017} In contrast, the active stresses result directly from cellular force generation.

If it were possible to separate the viscous stresses from the active, it would be possible to compare the viscous stresses directly to the kinematics, which would enable direct prediction of tissue flow for a given stress state. Unfortunately, such a clear separation is not possible without a measurement of the tissue viscosity, $\eta$. Common experimental approaches for measuring viscosity in passive systems, such as by applying a stress and quantifying the resulting flow, cannot be used here, because cell monolayers respond actively to forces, meaning that an experiment that applies an external stress to the cell monolayer will change the viscosity. Instead, experimental approaches must be devised to quantify tissue viscosity using the flows resulting from the cellular forces themselves.

One approach for studying the tissue viscosity comes from studying the dimensions of the friction constant $\xi$ and the tissue viscosity $\eta$. Dimensional analysis shows that the quantity $\lambda = \sqrt{\eta/\xi}$ has units of length, and is referred to as the hydrodynamic screening length. Conceptually, the hydrodynamic screening length defines the distance over which viscous stresses propagate throughout the monolayer, and, hence, it is one factor that is thought to affect the velocity correlation length.
However, other factors affect the velocity correlation length as well, such as the correlation lengths of stress and traction,\cite{tambe2011, vishwakarma2018, saraswathibhatla2021, dirusso2021, vazquez2022} meaning that $\lambda$ cannot be measured directly from studying the velocity data. 
One theory,\cite{alert2019prl} which was later supported by experiments,\cite{vazquez2022} proposed that $\lambda$ defines the distance between protrusions at the edge of an expanding cell layer. Hence, quantifying this distance is one method to quantify the hydrodynamic screening length $\lambda$, which in turn could reveal new insights about the cell-substrate friction $\xi$ or the tissue viscosity $\eta$.

A second approach to separate active from viscous forces is to compare the forces against the kinematics, namely the velocities and strain rates. For a (linear) passive system, the forces and kinematics are linearly proportional, but the active forces produced by the cells make the relationships between forces and kinematics complex. Hence, prior studies that have attempted to relate forces and kinematics have made little progress. For example, tractions sometimes align with and sometimes align against the velocity.\cite{trepat2009, kim2013, notbohm2016, bera2025} Additionally, across the entire cell monolayer, stresses are generally uncorrelated with strain rates.\cite{notbohm2016} While it would be ideal to measure tissue viscosity via an equation of the form $\eta = (\sigma - \sigma^a)/\dot{\varepsilon}$, such an approach is not possible because $\sigma^a$ is unknown. Because of this, alternate approaches to measure tissue viscosity must be established. Recently, we reported\cite{mccord2025arxiv} that the relationship between stress and strain rate varies over space, as follows. In regions of size a few cells to tens of cells, there is a clear linear correlation between shear stress $\sigma_s$ and shear strain rate $\dot{\varepsilon}_s$, with the slope being an effective viscosity $\eta^\mathrm{eff}$,  \textit{i.e.},  $\eta^\mathrm{eff} = \sigma_s / \dot{\varepsilon}_s$. Closer inspection of the effective viscosity is a second possible means of quantifying the tissue viscosity within a collectively migrating cell monolayer.

The objective of this manuscript is to quantify the tissue viscosity within a collectively migrating monolayer and to determine what factors increase or decrease it. As described above, we treat the monolayer as an active viscous fluid; in the Discussion section, we describe how our interpretation can be generalized to consider viscoelasticity. Given that there are no established methods to measure tissue viscosity, and that it is not possible to quantify the tissue viscosity directly from data of stress and strain rate, this manuscript attempts to measure the tissue viscosity using two approaches in parallel. The first approach uses the theoretical prediction\cite{alert2019prl} that the hydrodynamic screening length $\lambda$ defines the distance between protrusions at the edge of an expanding cell monolayer. The second studies the distribution of effective viscosity, computed by taking the ratio of shear stresses to shear strain rate at different locations within the monolayer. We began by studying the response to two chemical treatments: the first, CN03, activates Rho thereby increasing the presence of actin stress fibers and myosin activity;  the second, cytochalasin D, causes depolymerization of actin fibers. Both approaches for quantifying viscosity led to conclusions that are consistent across both treatments, namely, reducing the presence of actin fibers and myosin activity led to a reduction in the tissue viscosity, and increasing them led to an increase in the tissue viscosity. Next, we performed additional experiments to investigate further. Treatment with blebbistatin, a myosin II inhibitor, also led to a reduction in the tissue viscosity. Treatment with a metabolic inhibitor, which reduced activity without altering the cytoskeleton, did not alter the tissue viscosity. Finally, we found that reducing the cell-cell adhesions reduced the tissue viscosity.

\section{Results}

\subsection{Hydrodynamic Screening Length and Fluorescent Imaging as Indicators of Tissue Viscosity}

\subsubsection*{Morphology of the Edge of an Expanding Epithelial Cell Monolayer}

The first of our two approaches for quantifying tissue viscosity is based on a theoretical prediction that the hydrodynamic screening length $\lambda$ defines the distance between protrusions at the leading edge of an expanding cell monolayer,\cite{alert2019prl} as shown schematically in Fig. 1a. Although there exist other theories for formation of these protrusions, the focus has been on finger-like structures,\cite{sepulveda2013, mark2010, yang2020, kammeraat2025} which are more elongated than the protrusions present in our experiments. The elongated fingers typically form as result of flaws in a multicellular actin cable that forms at the edge of the cell monolayer.\cite{reffay2014} In a recent manuscript,\cite{vazquez2022}, we showed that the multicellular actin cable forms in response to how the cells were cultured before creating the free space for monolayer expansion. The typical approach is to culture cells against a barrier, remove the barrier, and allow the cell monolayer to expand. We recently showed that the multicellular actin cable forms over time against the barrier.\cite{vazquez2022} Our data showed that by culturing cells against the barrier for $<24$ hr, no actin cable is formed, meaning that the theory of Ref. \cite{alert2019prl}, which does not assume the presence of an actin cable, is relevant. We previously tested the theoretical prediction of Ref. \cite{alert2019prl}, and experimental results were consistent with the theory,\cite{vazquez2022} meaning that the distance between protrusions in an expanding cell monolayer is a straightforward means of measuring the hydrodynamic screening length $\lambda$. Given that $\lambda$ depends on the ratio of tissue viscosity and the cell-substrate friction, if one assumes that friction remains constant, if a perturbation changes the distance between protrusions, it would imply that the perturbation altered the tissue viscosity. Following this reasoning, we cultured Madin Darby canine kidney (MDCK) cells against a barrier and removed the barrier. Upon barrier removal, we verified that there was no multicellular actin cable at the edge of the monolayer (Supplemental Fig. S1). The cells were allowed to migrate for 24 hr, during which time protrusions formed at the edge of the cell monolayer (Fig. 1a,b). The protrusions in our experiments were shorter than the elongated finger-like structures reported previously\cite{poujade2007, petitjean2010, sepulveda2013, reffay2014}; this observation is expanded upon in the Discussion section.

\begin{figure}[th!] 
\centering
\includegraphics[keepaspectratio=true, width=6.5in]{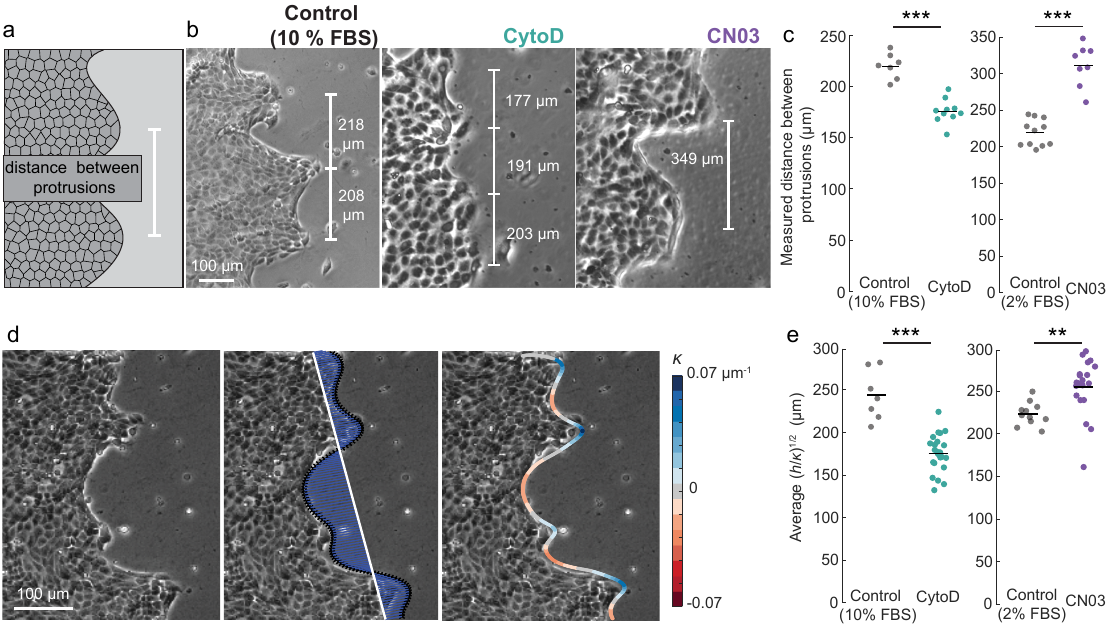}
\caption{Measurement of hydrodynamic screening length in a cell monolayer.
(a) Schematic of distance between protrusions, with the distance indicated by the white bar. 
(b) Phase contrast images of the leading edge of expanding cell monolayers in control conditions and treated with cytochalasin D or CN03. The protrusions are often led by cells having large lamellipodia. The manually measured distance between protrusions is shown.
(c) Summary of manually measured distance between protrusions for treatment with cytochalasin D ($p = 0.001$, two-sample t-test) and CNO3 ($p < 0.001$, two-sample t-test). Each dot represents the average distance between protrusions within a field of view, and black lines indicate means. 
(d) Quantification of average distance between protrusions. (Left) Representative image of an expanding cell monolayer. (Middle) Linear fit (white) and amplitude, $h$, (blue) of the leading edge. The amplitude is found by taking the orthogonal distance from the linear fit to the leading edge. (Right) Local curvature, $\kappa$, of the leading edge.
(e) Average of $\sqrt{h/\kappa}$ after 24 hr of migration for treatments with cytochalasin D ($p < 0.001$, two-sample t-test) and CN03 ($p = 0.004)$, two-sample t-test). Each dot represents an average over an independent field of view, and black lines indicate means. }
\label{fig1}
\end{figure}

\subsubsection*{Effect of Cytoskeleton on Hydrodynamic Screening Length}

Next, we quantified the distance between the protrusions at the 24 hr time point as a means of quantifying hydrodynamic screening length $\lambda$, which we use as an indicator of the tissue viscosity. Because viscosity reflects resistance to deformation, we reasoned that perturbations to the cytoskeleton would alter resistance to deformation and, hence, the tissue viscosity. We expected treatment with CN03, a Rho activator that increases stress fibers within each cell, to increase viscosity. Additionally, we expected treatment with cytochalasin D, which reduces the rate of F-actin polymerization, to lower the viscosity.
For different cell monolayers under the different conditions, the protrusions were identified visually by locating so-called leader cells, which had notable lamellipodia (Fig. 1b). The distance between adjacent protrusions was measured and averaged over each field of view. For cells treated with cytochalasin D, the distance between protrusions was smaller than for cells under control conditions (in 10\% serum, Fig. 1c).
Given that both CN03 and serum activate RhoA, the CN03 treatment was performed in medium with reduced fetal bovine serum (FBS) of 2\%, and the results showed that the distance between protrusions was greater for cells treated with CN03 compared to control (Fig. 1c).

To ensure that our measurements of distance between protrusions were unbiased, we repeated the measurement using an automated procedure, as in our prior work.\cite{vazquez2022} Briefly, the leading edge of the cell monolayer was identified with an edge detection algorithm, and then it was conceptually modeled as a straight line perturbed by a sine curve. Given that, for small perturbations, the second derivative of the sine curve is equal to the curvature, the wavelength is equal to the square root of amplitude divided by curvature. Following this idea, we quantified the curvature of the edge of the cell layer, $\kappa$, and the distance between a fitted straight line and the edge of the cell layer, $h$ (Fig. 1d). The distance between protrusions is then approximately given by $\sqrt{h/\kappa}$, which was quantified at each data point along the edge of an imaged cell island and averaged for each field of view. Compared to their respective controls, cytochalasin D and CN03 decreased and increased the value of $\sqrt{h/\kappa}$ (Fig. 1e), which is consistent with the trends in the manually measured distance between protrusions.

Given the prior studies\cite{alert2019prl, vazquez2022} showing that the distance between protrusions at the edge of the cell monolayer is equal to the hydrodynamic screening length $\lambda$, these data suggest that cytochalasin D decreased the hydrodynamic screening length and CN03 increased it. Hence, both treatments altered the balance between tissue viscosity and cell-substrate friction.

\subsubsection*{Fluorescent imaging of cytoskeleton and adhesions}

Next, we sought to identify how specifically the two treatments altered the hydrodynamic screening length $\lambda$. Given that $\lambda = \sqrt{\eta / \xi}$, it is possible that the treatments altered the tissue viscosity $\eta$, the substrate friction $\xi$, or both. To investigate this potential interplay, we fluorescently labeled the cytoskeleton, which produces forces, and the cell-cell and cell-substrate adhesions, which transmit the forces to neighboring cells and the substrate. Given that the viscosity represents a resistance to time-dependent changes in cell shape, we would expect that viscosity would be increased by a greater amount of actin in the cytoskeleton and/or a greater number of cell-cell adhesions. Additionally, studies have suggested that the cell-substrate friction $\xi$ increases with an increasing number of focal adhesions,\cite{garcia2015, ravasio2015, vazquez2022} meaning the number of focal adhesions is a semi-quantitative means of characterizing the friction.

In response to treatment with cytochalasin D, cells had notably fewer stress fibers, which is consistent with the fact that cytochalasin D reduces polymerization of F-actin (Fig. 2a and Supplemental Fig. S2). Additionally, cytochalasin D treatment reduced the relative intensity of E-cadherin at the cell-cell junctions (Fig. 2a, b). These two observations would suggest a reduction in viscosity caused by cytochalasin D. Cytochalasin D did not significantly alter the number of focal adhesions per cell, compared to control (Fig. 2a,c), which suggests no change in cell-substrate friction. Together, these observations suggest a reduction in tissue viscosity $\eta$ and no change in friction $\xi$, implying that the reduced hydryodynamic screening length caused by cytochalasin D (Fig. 1) is likely caused by a reduced tissue viscosity.

\begin{figure}[th!]
\centering
\includegraphics[keepaspectratio=true, width=6.5in]{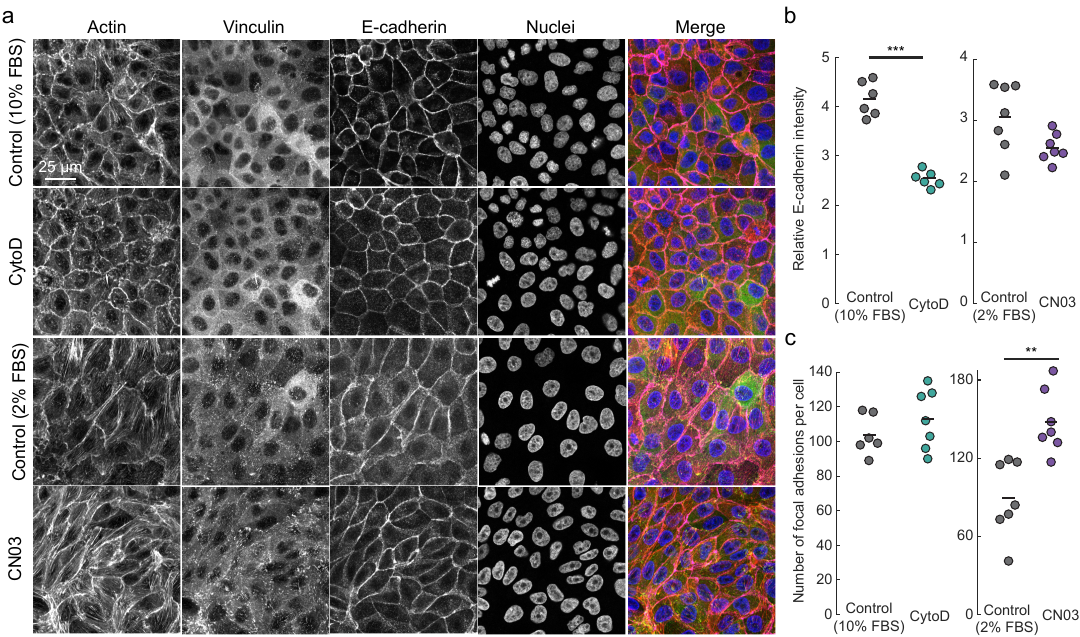}
\caption{Fluorescent labeling of the cytoskeleton, cell-cell, and cell-substrate adhesions. (a) Fluorescent images of actin, vinculin, E-cadherin, nuclei, and merged channels for control (10$\%$ FBS), cytochalasin D, control (2$\%$ FBS), and CN03 treated groups. (b) Relative E-cadherin intensity (ratio of fluorescent intensity at cell-cell junctions to intensity in cytoplasm) for treatment with cytochalasin D compared to its control ($p < 0.001$, two-sample t-test) and CN03 compared to its control ($p = 0.19$, two-sample t-test). (c) Average number of focal adhesions per cell  for treatment with cytochalasin D ($p = 0.18$, two-sample t-test) and CN03 ($p = 0.008$, two-sample t-test) compared to their respective controls. Each dot represents the average over a field of view, and black bars indicate means.}
\label{fig2}
\end{figure}

Compared to control, treatment with CN03 increased the number of focal adhesions (Fig. 2a, c), which would be expected to increase the friction $\xi$. Under this treatment, if the tissue viscosity $\eta$ were unchanged compared to control, the hydrodynamic screening length, $\lambda=\sqrt{\eta/\xi}$,  would have decreased. This is in contrast to observations in Fig. 1, which suggested that CN03 increased the hydrodynamic screening length. The only explanation for this discrepancy is that CN03 must have also increased the tissue viscosity. Turning to the images of actin, the CN03 substantially increased the number of actin fibers within the cells compared to control (Fig. 2a and Supplemental Fig. S2), which is consistent with prior observations.\cite{saraswathibhatla2020prx,saraswathibhatla_2022_PRE} Additionally, CN03 did not significantly change the amount of E-cadherin at the cell-cell junctions (Fig. 2a,b). These data suggest that CN03 increased the tissue viscosity by increasing the amount of fibrous actin within each cell. Additionally, the effects of increased actin were apparently far stronger than the effects of increased cell-substrate adhesions, implying the actomyosin cytoskeleton of the cell has a strong effect on the tissue viscosity.

In summary, the data from our first approach to study tissue viscosity suggest that cytochalasin D decreased the tissue viscosity whereas CN03 increased it. The data from both treatments suggest that a major factor affecting tissue viscosity is the number of actin stress fibers in the cytoskeleton. The treatment with cytochalasin D also suggest that the number of cell-cell adhesions may affect the tissue viscosity. However, this first approach is semi-quantitative and relies on interpreting the results through theoretical predictions on the effect of viscosity on hydrodynamic screening length. In the next section, we will address these limitations by introducing a second, more direct and quantitative approach to measure viscosity.

\subsection{Quantification of Viscosity by Ratio of Stress and Strain Rate}

\subsubsection*{Distribution of effective viscosity within the cell monolayer as an indicator of tissue viscosity and activity}

Next, we applied our second approach for quantifying tissue viscosity. Our method builds on our recent study,\cite{mccord2025arxiv} in which we quantified shear stress and shear strain rate, the ratio of which is an effective viscosity. We began by confining MDCK cells at a low density within 1 mm islands (Fig. 3a) and imaged over time to calculate the cell velocities, the strain rates, and the stress tensor within the plane of the cell monolayer. Our analysis will use the cell stresses and strain rates, which assumes a viscous behavior with negligible contributions from elasticity. To justify the choice to ignore potential elastic behavior, we ensured that the cell layer was within the fluid regime by quantifying the dimensionless shape index $q = < P/\sqrt{A} >$, where $P$ is the cell perimeter, $A$ is the cell area, and the brackets indicate a mean over all cells. If $q>3.81$, the monolayer is fluid-like.\cite{bi2015, park2015, bi2016} Our monolayers were in the fluid regime, with the average $q$ being 3.85 (Supplemental Fig. S3). 

\begin{figure}[th!]
\centering
\includegraphics[keepaspectratio=true, width=6.5in]{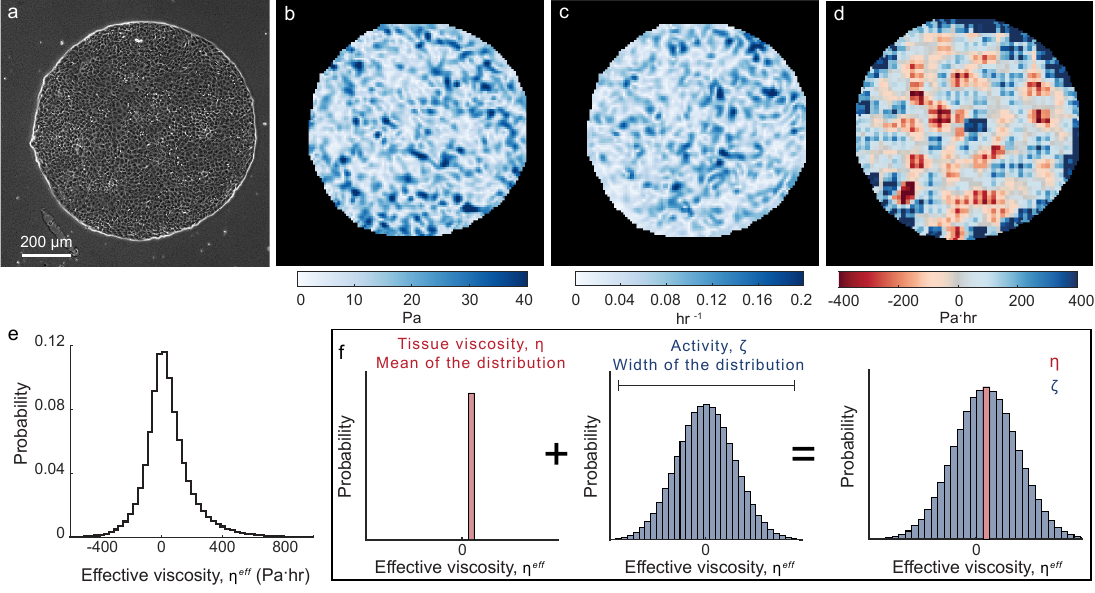}
\caption{Approach to measure effective viscosity within the monolayer.
(a) Phase contrast image of cell monolayer. 
(b, c) Color map of shear stress, $\sigma_s$ (b) and shear strain rate $\dot{\varepsilon}_s$ (c). 
(d) Color map of effective viscosity. 
(e) Histogram of effective viscosity across 6 cell islands over 15 hr. 
(f) Schematic showing how the distribution of effective viscosity is hypothesized to result from contributions of  tissue viscosity and activity. The tissue viscosity $\eta$ is hypothesized to have a small magnitude and positive mean. The active contribution $\zeta$ is hypothesized to have zero mean and a wide distribution, reflecting a wide range of activity in the monolayer. The sum is the effective viscosity $\eta^\mathrm{eff}$, which has a wide distribution and a slightly positive mean.
} 
\label{fig3}
\end{figure}

Ignoring bulk viscosity, the viscous stresses are connected to the cell flow through one parameter $\eta$ representing the tissue viscosity within the cell monolayer. As described in the Introduction, the experimentally measurable stresses are a combination of viscous and active, meaning it is not possible to quantify $\eta$ directly. Therefore, we study the \textit{effective} shear viscosity $\eta^\mathrm{eff}$ by quantifying the ratio between shear stress $\sigma_s = (\sigma_1-\sigma_2)/2$ and strain rate $\dot{\varepsilon}_s = (\dot{\varepsilon}_1-\dot{\varepsilon}_2)/2$ (Fig. 3b,c). Here, $\sigma_1$ and $\sigma_2$ are the first and second principal stresses, and $\dot{\varepsilon}_1$ and $\dot{\varepsilon}_2$ are the first and second principal strain rates. We use the term ``effective'' here to mean that $\eta^\mathrm{eff}$ results from a combination of the tissue viscosity and the activity produced by the cells. Heat maps of the shear stress and strain rate are shown in Fig. 3b and c, respectively. To account for spatial variation of $\eta^\mathrm{eff}$, we employ a windowing approach where the ratio of shear stress and strain rate is computed locally within a moving window across the monolayer. More detail, including information on the choice of window size, can be found in the Methods. In each window, we define $\eta^\mathrm{eff}$ as the ratio of shear stress and shear strain rate. By using this definition, we implicitly assume a linear viscous relationship. It is still unknown how accurate this assumption is, and it is possible that the relationship deviates from linear viscous, given that the rheology of the cytoskeleton is well described by a power law.\cite{fabry2001, fabry2003} Importantly, a linear viscous behavior typically fits well over a decade in time. A typical upper limit of shear strain rate in our experiments is 0.3 hr$^{-1}$, in which case the linear fit should be considered as reasonably accurate for strain rates in the range of 0.03--0.3 hr$^{-1}$. Indeed, representative scatter plots of shear stress against shear strain rate for different windows show reasonably linear trends (Supplemental Fig. S4).

The effective viscosity, $\eta^\mathrm{eff}$, was quantified for all widows in space and all points in time. A representative heat map of  $\eta^\mathrm{eff}$ at a single point in time is shown in Fig. 3d. 
In general, $\eta^\mathrm{eff}$ varies substantially over space and time, and, interestingly, depending on the orientation and magnitude of active stresses produced, it can even become negative, indicating injection of energy into the collective flow through shearing (shape changing) stresses.\cite{mccord2025arxiv} A histogram of $\eta^\mathrm{eff}$ was generated for six different cell islands at all points in space and over 15 hr of imaging. The histogram shows a wide distribution that is nearly symmetric and has a slightly positive mean (Fig. 3e). The distribution of $\eta^\mathrm{eff}$ in Fig. 3e results from a combination of the activity and the tissue viscosity. To demonstrate that the histogram of $\eta^\mathrm{eff}$ was not specific to MDCK cells, we also analyzed at the distribution of the effective viscosity over six HaCaT cell monolayers, and the main observations about the histograms were similar, namely the distribution of effective viscosity had a wide range with a mean that was positive (Supplemental Fig. S5).

We suggest here, as a working hypothesis, that coupled effects of tissue viscosity and activity on $\eta^\mathrm{eff}$ can be untangled by analyzing the histogram as follows. Thermodynamic considerations imply that the tissue viscosity, $\eta$, must be positive. $\eta$ could also vary over space and time, but it must always be positive. To avoid negative values of $\eta$, we expect the variation in $\eta$ to be no larger than its mean. Although $\eta$ represents a material property, it can be modulated by cellular activity through remodeling of the cytoskeleton and cell-cell adhesions,\cite{oriola2017} meaning that changes in activity can indirectly alter the tissue viscosity. The contribution to effective viscosity that results directly from activity, which we will call $\zeta$, can be positive or negative. Assuming randomness in cellular activity, we hypothesize that $\zeta$ has a mean near 0, meaning that, upon averaging over space and time, the same amount of energy is injected via active shearing stresses as is dissipated by active shearing stresses. Given our expectation that the variation in tissue viscosity $\eta$ is small and the observation that the width of the histogram of effective viscosity $\eta^\mathrm{eff}$ is large, we reason that the width of the histogram is due to the large variation in active viscosity. In summary, our working hypothesis is that the distribution of $\eta$ has a small, positive mean and negligible variability, which we express as $\eta \sim D(\eta,0)$, where $D$ represents a distribution with mean $\eta >0$ and zero variance. 
Additionally, we hypothesize that the distribution of $\zeta$ has a negligible mean and large variability, which we denote as $\zeta \sim D(0,S^2)$, meaning that the mean is zero and the variance $S^2$ is large. The distribution of effective viscosity is thus given by $\eta^\mathrm{eff} \sim D(\eta, S^2)$. 
This relationship is illustrated in Fig. 3f, where we schematically depict the expected distributions of $\eta$ and $\zeta$. As shown schematically in Fig. 3f, the histogram of $\eta^\mathrm{eff} = \eta + \zeta$ has a mean approximately equal to the mean of $\eta$ and a variability approximately equal to the variability of $\zeta$. Hence, our working hypothesis suggests that the tissue viscosity $\eta$ can be quantified by computing the mean value of the histogram in Fig. 3e.

\subsubsection*{Effective viscosity in response to perturbing actomyosin}

To test our working hypothesis that the mean and width of the histogram of $\eta^\mathrm{eff}$ serve as meaningful indicators of $\eta$ and $\zeta$, respectively, we measured the distributions of effective viscosity under the same treatments as used in Fig. 1. A shift in the mean of the effective viscosity distribution would suggest a change in the tissue viscosity, $\eta$, while a change in the width of the distribution would reflect alterations in active stress generation. 
In the paragraphs below, we study the distribution of $\eta^\mathrm{eff}$ in response to cytochalasin D and CN03 and interpret the data through the working hypothesis. 
If, following this approach, we arrive at a result that is consistent with the results in Figs. 1--2 or is otherwise known to be true, the results and reasoning together will provide evidence supporting the working hypothesis.

\begin{figure}[th!]
\centering
\includegraphics[keepaspectratio=true, width=5.5in]{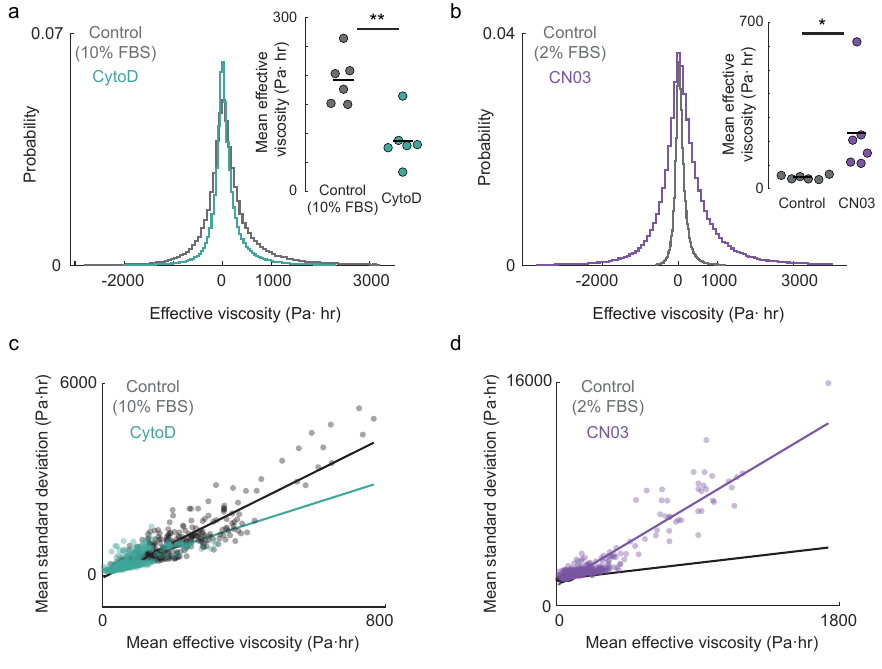}
\caption{Effect of altering actomyosin on effective viscosity.
(a) Distribution of effective viscosity for control ($10\%$ FBS) and treatment with cytochalasin D. Inset: Mean of effective viscosity for control ($10\%$ FBS) and treatment with cytochalasin D ($p = 0.002$, two-sample t-test). 
(b) Distribution of effective viscosity for control ($2\%$ FBS) and treatment with CN03. Inset:  mean effective viscosity for control ($2\%$ FBS) and treatment with CN03 ($p = 0.038$, two-sample t-test). 
In panels (a) and (b), each distribution is across 6 different cell islands per condition over 15 hr; each dot represents the mean over space and time for an independent cell island, and black bars indicate means over the dots. 
(c) Scatter plot of the standard deviation of effective viscosity against the mean effective viscosity for treatment with cytochalasin D and its control. The slope of the control was 5.52 and the slope with treatment with cytochalasin D was 3.52 ($p < 0.001$, analysis of covariance). 
(d) Scatter plot of the standard deviation of effective viscosity against the mean of effective viscosity for treatment with CN03 and its control. The slope of the control was 1.49 and the slope with treatment with CN03 was 7.58 ($p < 0.001$, analysis of covariance). 
In panels c and d, a dot represents the standard deviation and mean over space for one cell island at one point in time.}
\label{fig4}
\end{figure}

We began with the treatment of cytochalasin D, which, based on our interpretation of Figs. 1 \& 2, we expect to decrease the tissue viscosity. We seeded MDCK cells in 1 mm islands and imaged them for 1 hr, at which time we treated the cells with cytochalasin D or a vehicle control and imaged for an additional 14 hr. At each time point, we quantified the shear stresses and strain rates. Treatment with cytochalasin D caused a significant decrease in shear stress and shear strain rate, which is expected given that the treatment reduces the number of actin fibers and the cell contractility (Supplemental Fig. S6).  Compared to control, cytochalasin D also reduced the width of the distribution of effective viscosity ($p < 0.0001$, Levine's test for equality of variances, Fig. 4a). When interpreted through our working hypothesis, this observation implies that the treatment with cytochalasin D reduced the cell activity. Given that cytochalasin D decreases stress fibers and cell contractility, this is an expected result, hence it supports our working hypothesis. Turning now to the mean of the histogram, cytochalasin D also reduced the mean ($p = 0.001$, two-sample t-test, Fig. 4a), which, according to our working hypothesis, suggests a decrease in the tissue viscosity $\eta$. Hence, the data suggest that treatment with cytochalasin D reduced the tissue viscosity.

We then turned to the treatment of CN03, which, based on the results in Figs. 1 \& 2, we expect to increase the tissue viscosity $\eta$.  As above, 1 mm cell islands were imaged for 1 hr prior to treatment, then treated with CN03 or a vehicle control and imaged for an additional 14 hr, and the stresses and strain rates were quantified. Consistent with prior studies,\cite{saraswathibhatla2020prx, saraswathibhatla_2022_PRE, bera2025, mccord2025arxiv} the CN03 increased the shear stress and strain rate (Supplemental Fig. S6). The width of the histogram of effective viscosity was increased by the CN03 treatment ($p < 0.0001$, Levine's test for equality of variances, Fig. 4b). According to our working hypothesis, that the width of the histogram is an indicator of cell activity, this observation implies that CN03 caused greater activity, which is expected given that CN03 activates Rho, which in turn increases the number of actomyosin fibers within the cells. The mean of the histogram was also increased by the CN03 treatment ($p = 0.03$, two-sample t-test, Fig. 4b), which, according to our working hypothesis, would imply increased viscosity. This interpretation of the data is also consistent with our interpretation of the results in Figs. 1 \& 2.

It is notable that in the histograms of Fig. 4, the changes in mean and width coincide, that is, for CN03, both the mean and width increased, and for cytochalasin D, they both decreased. This raises the question of how closely the mean and width of effective viscosity are coupled. To explore this question, we computed the mean and standard deviation of effective viscosity for each cell island at every time point and plotted the mean against the standard deviation. We used the standard deviation as a proxy for the width of the effective viscosity distribution. The data showed a clear linear trend (Fig. 4c, d). Following our working hypothesis, this observation suggests that the tissue viscosity of the cell layer is proportional to its activity. There are two potential explanations for this relationship. First, as described above, the tissue viscosity could depend on the amount of actomyosin within the cytoskeleton. Similarly, activity depends on the amount of myosin within the cytoskeleton, meaning that the relationship between standard deviation and mean of effective viscosity observed here could result from a feedback inside the cytoskeleton that tends to balance the amounts of actin and myosin present. Secondly, activity could potentially change the tissue viscosity indirectly via changing the amount of cytoskeleton present\cite{oriola2017} or via nonlinear strain stiffening.\cite{mizuno2007}

A natural question that follows is how perturbations to the cytoskeleton affect the slope of the relationship between standard deviation and mean of effective viscosity. Given our interpretation that both the mean and standard deviation depend on the amount of actomyosin in the cytoskeleton, we would expect that increasing the amount of actomyosin by treatment with CN03 would increase the slope of the relationship between standard deviation and mean. Consistent with this expectation, the data show that, compared to control, the slope of standard deviation versus mean was significantly larger for cell islands treated with CN03 (Fig. 4d). By contrast, cytochalasin D had the opposite effect, with a slope that was significantly smaller than that for control conditions (Fig. 4c). Together, these results show that perturbations to the cytoskeleton consistently alter the mean and standard deviation of the effective viscosity in predictable ways, further supporting our interpretation that the mean of the effective viscosity distribution is the tissue viscosity, and the width of the distribution reflects the activity. 

\subsubsection*{Independent perturbation of cytoskeleton and activity}

So far, the inferences made by the two parallel approaches for measuring viscosity have been consistent. Specifically, the quantification of distance between protrusions (Fig. 1) coupled with the imaging of actin and adhesions (Fig. 2) suggest that CN03 increased the tissue viscosity and cytochalasin D decreased it. The distributions of effective viscosity (Fig. 4) suggested the same. Interestingly, the standard deviation and mean of the histogram, which we take as indicators of activity $\zeta$ and tissue viscosity $\eta$, respectively, were proportional (Fig. 4c,d), which raises the notion that activity may affect tissue viscosity, as suggested previously.\cite{oriola2017, blanch2017} It is not clear whether activity is the only factor affecting viscosity. To investigate this possibility, we sought to perturb the cytoskeleton and activity independently.

To begin, we sought to perturb the active contribution of effective viscosity, $\zeta$, without affecting the cytoskeleton. To this end, we inhibited metabolism directly, expecting that metabolic inhibition would reduce the activity without affecting the cytoskeleton, as depletion of ATP occurs quickly, while reorganization of the cytoskeleton occurs over longer timescales.\cite{sutton2001} The cells were imaged over time before and after inhibiting oxidative phosphorylation and glycolysis by treating with both sodium cyanide (NaCN) and 2-Deoxy-d-glucose (2-DG). To ensure that the cells remained viable after the treatment, we imaged them for a relatively short time period of 50 min after treatment. During this time frame, the average shear stresses and strain rates were reduced in response to the metabolic inhibition, but they remained nonzero, being consistently above 5 Pa and 0.01 hr$^{-1}$, respectively (Supplemental Fig. S7), indicating the cells remained alive. Compared to control, the metabolic inhibition significantly reduced the width of the histogram of effective viscosity (Levene's test for equality of variances $p < 0.001$, Fig. 5a). The standard deviation of effective viscosity decreased from 746 Pa$\cdot$hr in the control condition to 496 Pa$\cdot$hr following metabolic inhibition. 

\begin{figure}[b!]
\centering
\includegraphics[keepaspectratio=true, width=6in]{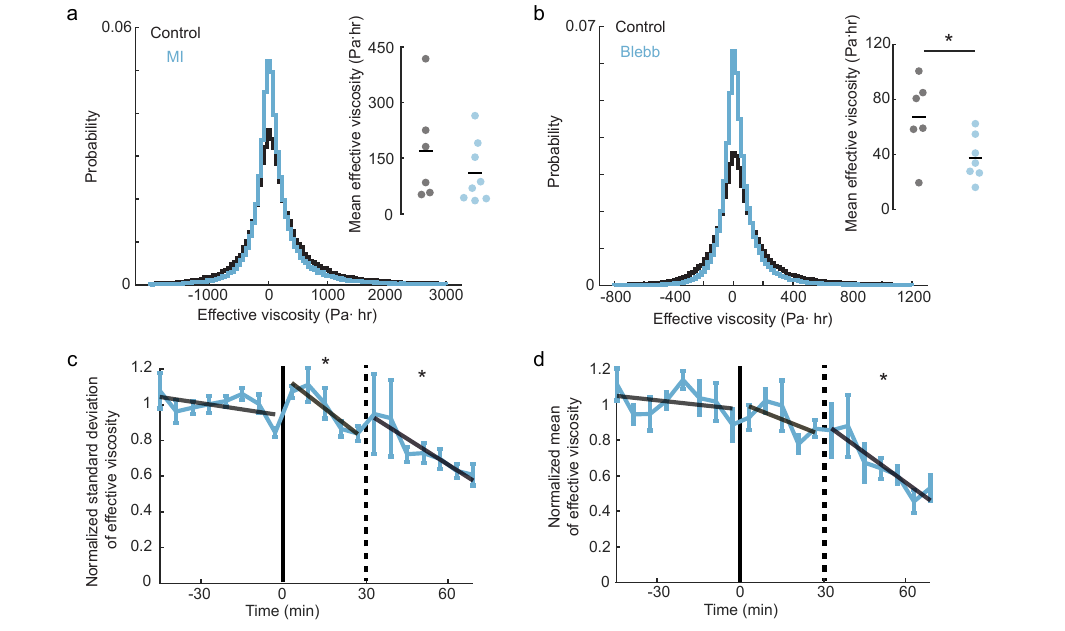}
\caption{
The width of the histogram of effective viscosity is controlled by cellular activity.
(a) Histograms showing effective viscosity distributions for control (gray) and metabolic inhibition (blue) groups. Data shown is from 6 control and 8 metabolically inhibited  cell islands over 1 hr of imaging. Levene's test indicated a statistically significant difference in the variance between control and metabolic inhibition ($p < 0.0001$). Inset: Mean effective viscosity for control and metabolic inhibition groups ($p = 0.343$, two-sample t-test). 
(b) Histograms showing effective viscosity distributions for control (gray) and blebbistatin (blue) groups. Data shown is from 6 control and 7 blebbistatin-treated cell islands over 2 hr of imaging. Levene's tested indicated a significant difference in the variance between control and blebbistatin groups ($p < 0.0001$). Inset: Mean effective viscosity for control and blebbistain groups ($p = 0.03$, two-sample t-test). 
In panels a and b, each dot represents a different cell island; black bars indicate means. 
(c, d) The normalized standard deviation (c) and mean (d) of the distribution of effective viscosity over time before and after treatment with blebbistatin. Data for each cell island were normalized to the values before treatment ($t<0$). The treatment occurred at $t=0$. The dashed lines represent the first 30 min post treatment. Lines represent averages over all 7 blebbistatin-treated cell islands, and bars represent the standard error of the mean.
For panel c, the slopes and 95$\%$ confidence intervals were -0.141 [-0.401, 0.118] min$^{-1}$ for time $\in$ (-50, 0) min, -0.722  [-1.27, -0.169] min$^{-1}$ for time $\in$ (0, 30) min, and -0.591 [-0.839, -0.344] min$^{-1}$ for time $\in$ (30, 70) min.
For panel d, the slopes and 95$\%$ confidence intervals were -0.0987 [-0.427, 0.23] min$^{-1}$ for time $\in$ (-50, 0) min, -0.361 [-1.27, 0.547] min$^{-1}$ for time $\in$ (0, 30) min, and -0.663 [-0.972, -0.354] min$^{-1}$ for time $\in$ (30, 70) min.
In panels c and d, the * symbol represents time periods for which confidence intervals did not span 0, indicating slopes that were statistically different from 0.
} 
\end{figure}

Despite the reduction in the width of the histogram, the mean effective viscosity remained statistically unchanged (Fig. 5a, inset; $p = 0.34$, two-sample t-test), which, following our working hypothesis, we interpret to indicate that the tissue viscosity remained unchanged. 
For further evidence, we considered that the tissue viscosity could depend on the cytoskeleton (as suggested above) and cell-cell adhesions (which are studied in detail in the next section). Given that the data suggest no change in tissue viscosity, we would expect no change in the cytoskeleton or adhesions. Consistent with this expectation, fluorescent imaging of actin stress fibers, focal adhesions, and junctional E-cadherin 50 min post treatment showed no discernible differences in cytoskeleton organization or adhesion structures between control and metabolically inhibited conditions (Supplemental Fig. S8). Hence, the data indicate that metabolic inhibition primarily reduced activity. One caveat to this approach is that perturbations to metabolism have only modest effects,\cite{mccord2025arxiv} meaning we cannot rule out the notion that a larger perturbation to activity could affect viscosity indirectly, for example by activity-driven changes to the cytoskeleton\cite{oriola2017} or activity-induced stiffening of the cytoskeleton.\cite{mizuno2007, wang2002}

We next sought to perturb the activity $\zeta$ independently from the tissue viscosity $\eta$ with yet another approach. For this we used blebbistatin, which inhibits cell contraction. Importantly, blebbistatin affects the cell differently than cytochalasin D. Blebbistatin does not directly affect polymerization of actin fibers; instead, it blocks myosin II from contracting. As a result, it acts quickly, within $\approx$2-3 min.\cite{goeckeler2008} At later time points, $\approx$30 min after blebbistatin treatment, the cells respond to the treatment by remodeling their cytoskeleton, reducing the amount of fibrous actin present.\cite{goeckeler2008} 
Hence, we hypothesized that blebbistatin would have two effects. First, at time points immediately after treating with blebbistatin, cells would exhibit reduced activity, meaning the distribution of effective viscosity $\eta^\mathrm{eff}$ would become narrower, but its mean would remain nearly constant. Then, at later time points, the reduction of actin caused by cytoskeletal remodeling would also cause a reduction in the mean of the distribution of $\eta^\mathrm{eff}$, reflecting a reduced tissue viscosity $\eta$. 
Following this hypothesis, we imaged cell islands for 50 min, treated with 20 {\textmu}M blebbistatin, and then imaged for an additional 70 min. From these data, we normalized the standard deviation and mean of $\eta^\mathrm{eff}$ to their values before blebbistatin treatment, and then plotted the data over time (Fig. 5c,d). As an indicator of whether or not the blebbistatin treatment had an effect on the standard deviation or mean of $\eta^\mathrm{eff}$, we computed the slopes, \textit{i.e.}, the change in standard deviation or mean over time.
During the first 30 min after treatment, the standard deviation of $\eta^\mathrm{eff}$ had a 95\% confidence interval of the slope that was negative, indicating a statistically significant reduction in activity, as we hypothesized (Fig. 5c). 
The change in time of the mean, however, had a 95\% confidence interval that spanned 0 in the first 30 min after treatment, suggesting no change in tissue viscosity during these early time points (Fig. 5d). For later times, both the standard deviation and the mean of the distribution of $\eta^\mathrm{eff}$ had a statistically significant negative trend in time (Fig. 5c,d). 

Direct inspection of the histogram of effective viscosity for times $>30$ min after blebbistatin treatment indicate that blebbistatin reduced both the standard deviation and the mean, indicating reductions in both activity and tissue viscosity occurred at late time points (Fig. 5b). 
The reduction in mean effective viscosity was approximately a factor of 2, which occurred with a factor of $\approx 2$ change in shear stress (Supplemental Fig. S7). In comparison, a prior study that inferred viscosity by fitting experimental data against a theory found that reducing the stress by a factor of $\approx$10 by treatment with blebbistatin led to a similar factor of $\approx$10 reduction in viscosity.\cite{blanch2017} 

In summary, the data show that at early time points, blebbistatin decreased activity with no statistically significant change in the tissue viscosity. At later time points, the cells had time to remodel their cytoskeleton in response to the blebbistatin treatment, meaning both activity and the amount of F-actin in the cytoskeleton were reduced, and in response the tissue viscosity decreased. 

\subsubsection*{Effective viscosity in response to perturbing cell-cell adhesions}

\begin{figure}[th!]
\centering
\includegraphics[keepaspectratio=true, width=6.3in]{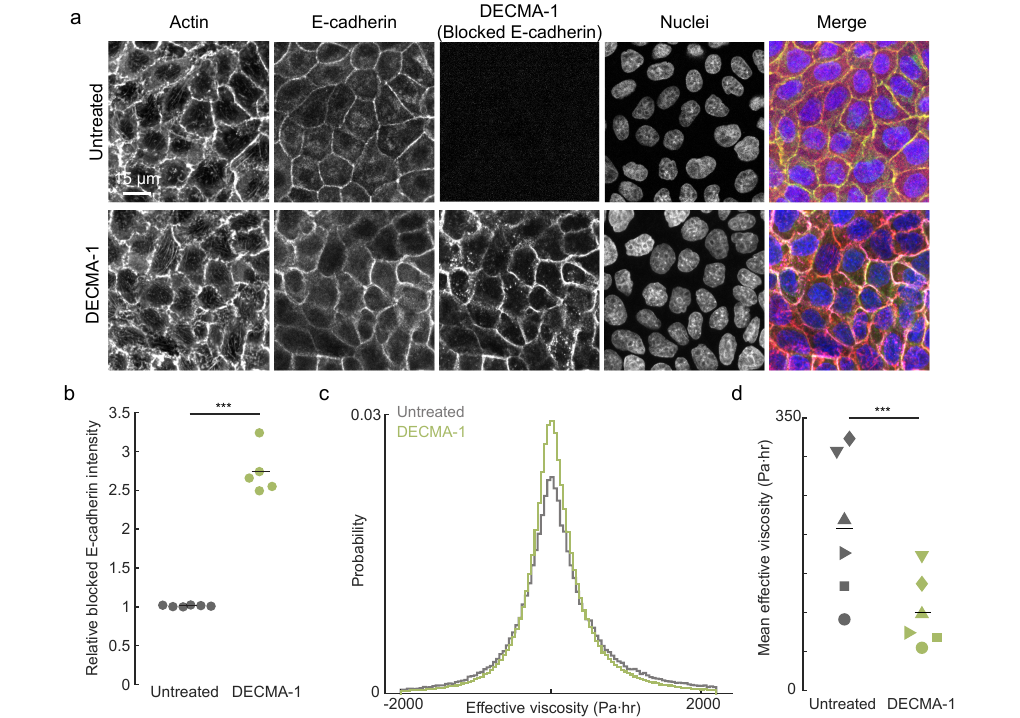}
\caption{Effect of cell-cell adhesion on tissue viscosity. 
(a) Fluorescent images of actin, E-cadherin, E-cadherin blocked by DECMA-1, nuclei, and merged channels for pre-treatment and DEMCA-1 treated groups. 
(b) Relative fluorescent intensity of blocked E-cadherin (ratio of intensity at cell-cell junctions to intensity in cytoplasm) for pre-treatment and DEMCA-1 treated groups  ($p < 0.001$, two-sample t-test). Each dot represents the average over a field of view, and black bars indicate means.
(c) Distribution of effective viscosity for pre-treatment and DECMA-1 treated groups ($p = 0.32$, Levine's test for equality of variances). 
(d) Mean effective viscosity for pre-treatment and DECMA-1 treated groups ($p = 0.004$, one-sample t-test). Each marker represents the average over space and time for an independent cell island, and black bars indicate means. The marker shape is the same cell island before and after treatment with DECMA-1.} 
\end{figure}

Next we considered one more cellular structure that might affect the tissue viscosity, namely cell-cell adhesions. Given that cell-cell adhesions could potentially support forces that resist cellular rearrangements within the monolayer, we hypothesized that reducing adhesion at the cell-cell interface would decrease the tissue viscosity, $\eta$. To test this hypothesis, we treated our monolayers with DECMA-1, an E-cadherin-blocking antibody that binds to extracellular E-cadherin, thereby reducing cadherin-based cell-cell adhesions.\cite{choi2022} 1 mm cell islands were treated with 5 $\upmu$g/mL DECMA-1 E-cadherin antibody and the islands were fixed after an additional  17 hr. We first confirmed antibody engagement and junctional adhesion disruption through fluorescent labeling of the cytoskeleton, E-cadherin, and blocked E-cadherin for samples with and without DECMA-1 treatment. Images of actin showed no changes in cytoskeletal structure between pre- and post- treatment conditions (Fig. 6a). As an indication of the amount of E-cadherin blocked by DECMA-1, we fluorescently labeled and imaged the DECMA-1 itself, which was not present in pre-treatment conditions but was present post-treatment (Fig. 6a). To quantify these results, we calculated the ratio of junctional to cytoplasmic blocked E-cadherin. In pre-treatment controls, the ratio was near 1, indicating random noise, and as expected, DECMA-1 treated samples had a significantly higher ratio (Fig. 6b), indicating that DECMA-1 localized to the cell-cell adhesions.

Next, the cadherin blocking procedure was repeated in experiments that measured shear stresses and strain rates. Treatment with DECMA-1 did not have a significant effect on shear stress or shear strain rate (Supplemental Fig. S6). Additionally, the treatment did not have a statistically significant effect on the width of the distribution of effective viscosity (Fig. 6c). When interpreted through our working hypothesis, this observation implies that treatment with DECMA-1 did not alter the cell activity, which is consistent with the observations that DECMA-1 did not change the cytoskeleton (Fig. 6a) or the shear stress (Supplemental Fig. S6). Looking now at the mean of the distribution of effective viscosity, DECMA-1 treatment significantly reduced the mean (Fig. 6d), which according to our working hypothesis, implies reduced tissue viscosity. These results indicate that blocking cell-cell adhesions reduced the tissue viscosity and demonstrate a situation wherein tissue viscosity was changed with no change in activity.

\section{Discussion}

Although both motion and force can be measured during collective cell migration, direct comparison between the two has remained a challenge, because the forces are a combination of active and viscous components. The objective of this study was to quantify the tissue viscosity as a means of separating the active from viscous components of the stress tensor within a cell monolayer. To this end, we employed two complementary approaches. The first, using distance between protrusions at the edge of the cell monolayer, suggested that increasing and decreasing the amount of actin in the cytoskeleton respectively increased and decreased the tissue viscosity. 
The second compared shear stresses and shear strain rates in local regions of space within the monolayer. Results were consistent with the first method, namely increasing (decreasing) the presence of actin stress fibers in the cytoskeleton increased (decreased) the tissue viscosity. In subsequent experiments, we also showed that decreasing cell-cell adhesion decreased the tissue viscosity. These findings have two major implications. First, we present two, distinct, experimentally accessible methods for estimating viscosity within the monolayer. Second, and more fundamentally, we demonstrate that the cytoskeleton and cell-cell adhesions strongly determine the effective viscosity, with perturbations to either component producing predictable changes in tissue viscosity, thereby establishing a link between subcellular structure and macroscopic material properties of epithelial tissues. 

Within this system, inertial effects are negligible, meaning that force and motion must be related through a viscous relationship. It is typically assumed that the viscosity linearly relates stress and strain rate, \cite{marchetti2013review, alert2020review} but until now this assumption has been untested, and the value of viscosity has been uncertain. Order-of-magnitude estimates have suggested that viscosity in cell monolayers lies within a broad range---from $\sim$100 Pa$\cdot$hr to $\sim$10,000 Pa$\cdot$hr.\cite{duclos2017, perez-gonzalez2019} A similar range of values has been inferred by fitting models to experimental data.\cite{marmottant2009, guevorkian2010, blanch2017} Our results showed that tissue viscosity is approximately in the range of 50--250 Pa$\cdot$hr. Perturbing cell-cell adhesions or actin polymerization in the cytoskeleton reduced the tissue viscosity by a factor of 2, whereas activating actomyosin in the cytoskeleton increased tissue viscosity by nearly a factor of 5 (Fig. 4). 
Given that the magnitude of tissue viscosity has been unknown until now, the value of hydrodynamic screening length $\lambda$ has also been unclear. Order-of-magnitude estimates have suggested that $\lambda$ ranges from tens to hundreds of microns.\cite{cochet2014, duclos2017, duclos2018, alert2019langmuir, perez-gonzalez_2019} Our results show that for MDCK cells $\lambda$ is $\approx$230 {\textmu}m for control conditions in 10\% serum, which is approximately double the value reported for a different cell type, HaCaT.\cite{vazquez2022} Our data also suggest that for MDCK cells, the value of $\lambda$ can be reduced or increased by a factor of $\approx$1/3 by perturbations that reduce or increase the tissue viscosity.

The quantification of hydrodynamic screening length $\lambda$ relied on a theoretical prediction\cite{alert2019prl} that the distance between protrusions at the leading edge of an expanding cell monolayer is proportional to $\lambda$. The theory treated the cell monolayer as an active fluid, with propulsive tractions at the edge of the cell layer that pull the cells into free space. Following this assumption, cells at the edge tend to move faster than the overall rate of monolayer expansion. For the monolayer to remain confluent (with no holes forming), some cells at the edge of the monolayer must migrate more slowly such that the average speed at the edge is matched by the average speed within the bulk. As a result, the edge of the cell layer becomes rough with protrusions forming, as in our experiments. The theory predicts that the characteristic size of a protrusion at the edge is equal to the hydrodynamic screening length $\lambda$. A major difference between this theory and others\cite{sepulveda2013, mark2010, yang2020, kammeraat2025} is that it is missing two features that can sometimes occur in monolayer expansion experiments. The first is a multicellular actin cable, which smooths the edge of the cell monolayer. The second is leader cells, which form at local flaws in the multicellular actin cable.\cite{reffay2014} Under these conditions, the monolayer forms long finger-like structures, often tens of cells in length and only a few cells wide.\cite{poujade2007, petitjean2010, sepulveda2013, reffay2014} It was proposed\cite{yang2020} that an essential factor for forming the large, elongated fingers is for leader cells (such as those that form in response to a flaw in the actin cable\cite{reffay2014}) and the followers to be phenotypically different, with elevated motility. It was further proposed by Ref. \cite{yang2020} that, because the theory of Ref. \cite{alert2019prl} does not consider leaders and followers, the structures that form at the edge would be more stable than the long fingers observed elsewhere. Our data is consistent with this idea, as the protrusions in our study were not elongated fingers; instead, they were short, with a small aspect ratio of $\approx 1$ (Fig. 1). Given that there continue to be new explanations posed as to how fingers form in an expanding cell monolayer\cite{kammeraat2025} it cannot be ruled out that the theory we use is incorrect, in which case the relationship between curvature and viscosity in our experiments is a coincidence. We emphasize, however, that it is striking that inferences made about effects of both cytochalasin D and CN03 on viscosity match the inferences of our second method. Further notable is that a prior study of ours used the same method to infer a relationship between substrate stiffness and hydrodynamic screening length\cite{vazquez2022} and that the small aspect ratio of protrusions in our study matches inferences made by Ref. \cite{yang2020}.

At first glance, it may seem that the two approaches for quantifying tissue viscosity used here might be inconsistent: the first approach assumed that the tissue viscosity is constant over space, and the second approach shows notable spatial variation. However, this apparent contradiction is resolved by interpreting the distribution of effective viscosity through our working hypothesis, which is that the tissue viscosity has a small, positive mean and small variability, whereas the active viscosity has negligible mean and large variability. We initially posed this hypothesis based on the observation that the histogram of effective viscosity had a wide range, with many values even being negative, but it always had a small, positive mean value. We tested this hypothesis with several experiments, with all findings consistent with the  hypothesis. Hence, the hypothesis, together with the supporting data, suggest that spatial variation in the tissue viscosity is likely to be relatively small, which matches the assumptions made by the first approach for quantifying tissue viscosity. Hence, the two methods for quantifying tissue viscosity are consistent and complementary: both approaches offer an estimate of average the average tissue viscosity, and the second approach additionally quantifies how activity produces local deviations from viscous behavior. 

Because the time scales for cell motion (tens of minutes to hours) are far larger than those for turnover of actin and E-cadherin (tens of seconds to minutes),\cite{khalilgharibi2019, debeco2009} our study treated the monolayer as being a fluid. This assumption can be relaxed by considering the monolayer to be viscoelastic. In the simplest model for a viscoelastic fluid, the Maxwell model, the viscous term $\eta$ is related to the elastic term $E$ through a characteristic relaxation time $\tau$ according to $\eta = E\tau$. Hence, the perturbations in our experiments that have increased and decreased the tissue viscosity $\eta$ at long time scales (tens of minutes) would be expected to increase and decrease the elastic modulus at time scales shorter than the $\sim$1 min required for turnover of actin and E-cadherin. Following this reasoning, it would be expected that cytochalasin D, blebbistatin and CN03, which respectively decreased, decreased, and increased tissue viscosity on long time scales would decrease, decrease and increase the elastic modulus on short time scales. These results are intuitive and self-consistent, meaning that the tissue viscosity $\eta$ can be thought of equivalently as the short term elastic modulus $E$ multiplied by a time scale for relaxation. This interpretation enables us to apply prior findings about the elastic modulus of the cytoskeleton to infer the implications for viscosity of the epithelial tissue. For example, the stiffness of actin networks can be strongly increased by myosin activity,\cite{mizuno2007} and the stiffness of the cytoskeleton is approximately linearly proportional to cell force production.\cite{wang2002} Hence, we would predict that a sufficiently large increase in myosin motor activity would cause greater tissue viscosity, which is not inconsistent with the inference from Fig. 4 that the tissue viscosity is approximately linearly correlated to activity.

A central goal of this study is to measure the tissue viscosity, which is a material property. In materials science, it is common to link material properties to microstructural features such as crystal structure, defects, and grain size in metals and ceramics or architecture and crosslinking in polymers and fibrous materials. Our study takes an analogous approach by revealing the microstructural features that affect tissue viscosity. One factor affecting the tissue viscosity is the cytoskeleton---increasing actomyosin through CN03 or decreasing actomyosin with cytochalasin D led to a corresponding increase or decrease in tissue viscosity (Fig. 4a,b). A second factor is the cell-cell adhesions, whereby blocking E-cadherin caused a reduction in tissue viscosity (Fig. 6). Thirdly, activity within the cytoskeleton potentially contributes to the tissue viscosity---although results of Fig. 5 cannot conclusively indicate that activity affects viscosity, the data in Fig. 4c,d are consistent with this notion. These results establish a direct relationship between the structure and activity of the cell and the material properties of the tissue. Through these experiments, we have shown that the concept of viscosity, which is typically reserved for passive materials, can be meaningfully extended to describe the mechanical behavior of active biological systems.

Although the methods to quantify forces (both cell-substrate tractions and intercellular stresses) during collective cell migration were established over a decade ago, it has been challenging to use experimental measurements of tractions and stresses to test theoretical assumptions and to quantify model parameters, because the forces are a combination of active and viscous. By quantifying the tissue viscosity, our study provides a new path forward to test theoretical assumptions and provides experimentally grounded estimates of the tissue viscosity across multiple perturbations. Our experimental approach will lead to new avenues of research that clarify the biophysical meaning of the different terms in theories and models for collective migration, improving their predictive capability. Beyond theory, our results provide new insight on the tissue's resistance to flow: increasing (decreasing) the cytoskeleton or cell-cell adhesion increases (decreases) the tissue's resistance to flow. This finding raises new questions as to how viscosity modulates, and is modulated by, other physical and biological behaviors of epithelial tissues such as the fluid--solid transition and the epithelial--mesenchymal transition. Our study provides a starting point to investigate these questions.

\section{Methods}

\subsection{Cell culture}
Madin-Darby Canine Kidney (MDCK) type II cells were maintained in 1 g/L glucose Dulbecco's Modified Eagle's Medium (DMEM, 10-014, Corning) supplemented with 10$\%$ FBS (Corning) and 1$\%$ Penicillin-Streptomycin (Corning). Human keratinocytes (HaCaTs) were maintained in 4.5 g/L glucose Dulbecco's Modified Eagle's Medium (DMEM, 10-013, Corning) supplemented with 10$\%$ fetal bovine serum (FBS, Corning) and 1$\%$ Penicillin-Streptomycin (Corning). All cells were maintained at 37$^\circ$C and 5$\%$ CO$_2$. Unless otherwise stated, cells were moved into medium containing 2$\%$ FBS 8 hr prior to the beginning of the experiment.

\subsection{Chemical treatments}
Cells were treated with 20 {\textmu}M blebbistatin (Sigma-Aldrich), 0.05 {\textmu}M cytochalasin D (Sigma), or 2 {\textmu}g/mL CN03 (Cytoskeleton Inc.). Blebbistatin and cytochalasin D were dissolved in DMSO, and CN03 was dissolved in water. The stock solutions were then diluted in phosphate buffered saline (PBS) to obtain the desired concentrations for experiments. 

\subsection{Substrate preparation and time-lapse imaging}
For experiments involving cell islands, cells were confined in 1 mm diameter circles on polyacrylamide gels of a 6 kPa Young's modulus embedded with florescent particles using previously described methods.\cite{notbohm2016, saraswathibhatla2020sciData} Briefly, 1 mm circular holes were punched into polydimethlysiloxane (PDMS) sheets using a biopsy punch. The sheets were then adhered to the polyacrylamide gels. The covalent crosslinker sulfo-SANPAH (50 mg/mL) was used to coat the holes with 0.1 mg/mL type I rat tail collagen (Corning). Cells were then seeded onto the masks, allowed to adhere for 1 hr, and the masks were removed. Cells were allowed to come to confluence overnight. For control experiments and experiments involving treatment with cytochalasin D and CN03, cell islands and fluorescent particles were imaged every 15 min for 15 hr. Cell islands and fluorescent particles were imaged every 10 min for 1 hr and every 6 min for 1.6 hr for metabolic inhibition and blebbistatin experiments, respectively. For DECMA-1 experiments, cells islands and fluorescent particles were imaged every 15 min for 20 hr. After imaging, cells were removed by incubating in 0.05 $\%$ trypsin, and images of a traction-free reference state were collected. 

For experiments involving distance between protrusions, wound healing experiments were carried out as described previously.\cite{vazquez2022} Briefly, barriers were prepared by mixing -200 mesh iron powder into PDMS, and cured at as a sheet. Strips were cut from the sheet and were spin coated in an additional layer of PDMS to prevent oxidation of the iron filings. The strips were then soaked overnight in 2$\%$ Pluronic F-127 (Sigma). After coating the surface of the gel in collagen, barriers were placed on the gel and held in place with a magnet seated below the glass bottom dish. Cells were seeded in a confluent monolayer next to the barrier, allowed to adhere for 1 hr, and the barrier was removed by taking the dish off the magnet, and placing a second magnet over the barrier until the barrier lifted from the gel. Cells were then allowed to migrate into the free space, and phase contrast images were taken after 24 hr of migration. 

All live cell imaging was performed at 37$^\circ$C and 5$\%$ CO$_2$ using a custom built incubator cage. An Eclipse Ti microscope (Nikon Instruments) was used with a 4$\times$ or 10$\times$ objective and an Orca Flash 4.0 digital camera (Hamamatsu) running Elements Ar software (Nikon).

\subsection{Quantification of curvature and amplitude along the edge of an expanding monolayer}

To create an expanding cell monolayer, cells were seeded against a barrier, allowed to adhere for 1 hr, and the barrier was removed. The 1 hr time point was chosen to prevent the formation of a multicellular actin cable at the leading edge of the monolayer,\cite{vazquez2022} (Supplemental Fig. S1) as flaws in the multicellular actin cable lead to formation of leader cells.\cite{reffay2014,begnaud2016} As described in the Results section, the theoretical prediction\cite{alert2019prl} that we use to relate morphology of the monolayer edge to hydrodynamic screening length $\lambda$ assumes no multicellular actin cable and no leader cells, making it appropriate for our experiments. Chemical treatments were added at the time of barrier removal, and the cells were allowed to migrate into the free space for 24 hr.

Curvature and amplitude along the leading edge were computed using methods described previously. \cite{vazquez2022} Briefly, phase contrast images of the leading edge from the 24 hr time point were binarized, and boundaries were identified using a Sobel edge detection filter, following previously established methods.\cite{treloar2013} Detected edges were then dilated and filtered to remove small or spurious region, and the coordinates of the boundary of the cell monolayer were identified. These coordinates were then used to compute the distance along the boundary and the angle tangent to the boundary, from which the curvature was computed using a linear regression of angle against distance, with the slope being the curvature, $\kappa$. The local height, $h$ of the monolayer edge was computed by fitting a line to the boundary, and computing the distance from the fitted line to the boundary via orthogonal projection. The root-mean-square of $\kappa$ and $h$ were computed for each field of view. 

The distance between protrusions was measured manually in ImageJ. Protrusions were identified by locating peaks where leader cells had visible lamellipodia, and the distance between adjacent peaks was measured. These distances were then averaged over each field of view. 

\subsection{Fixed cell imaging and quantification of focal adhesions and relative E-cadherin intensity}
For cytoskeletal imaging, experiments were carried out as described above, and at the time point of interest, cells were fixed and stained. To fix the cells, cells were rinsed twice with PBS and fixed with 4\% paraformaldehyde for 20 min. The cells were then washed twice with PBS and permeabilized with 0.1\% Triton X-100. Actin staining was carried out using ActinRed 555 ReadyProbe Reagent (ThermoFisher R37112) per manufacturer instructions. Focal adhesions and E-cadherin was stained overnight using Vinculin Monoclonal Antibody (7F9), Alexa Fluor 488, (ThermoFisher 53-9777-82, 1:250 dilution) and CD324 (E-cadherin) Monoclonal Antibody (DECMA-1), eFluor 660 (ThermoFisher 50-3249-82, 1:250 dilution), respectively. Cell nuclei were stained using NucBlue Fixed Cell Stain (ThermoFisher R37606) per manufacturer instructions.  

Imaging of fixed samples was performed on a Nikon A1R confocal microscope (Nikon Instruments) running NIS-Elements Ar software (Nikon). Imaging was performed using a 40$\times$ numerical aperture 1.15 water immersion objective, and images were taken from the base to the apex of the cells. To visualize E-cadherin, cortical actin, and nuclei, a maximum intensity projection through the cell height was computed. For visualization of stress fibers and focal adhesions a maximum intensity projection of the confocal slices near the base of the cells was computed. Focal adhesions per cell were quantified using images of vinculin. First, the images were binarized, and regions within a defined area range ($\sim$ 0.5--2 {\textmu}m$^2$) were retained, consistent with the expected size of focal adhesions in MDCK cells ($\sim$ 1{\textmu}m$^2$) \cite{kuo2014}. Cell boundaries were computed with SeedWater Segmenter, and the number of focal adhesions within each cell was quantified. This was then averaged over all cells within the field of view. To quantify relative E-cadherin intensity, the ratio of E-cadherin fluorescent intensity at the cell boundary to the cytoplasm was computed and averaged over each field of view. 

\subsection{Cell segmentation}
Phase contrast images of cells were segmented using CellPose (cyto2 model).\cite{stringer2021}Cytoskeleton images were segmented using SeedWater Segmenter.\cite{mashburn2012}

\subsection{Image correlation, traction force microscopy, and monolayer stress microscopy}
Cell-induced substrate displacements were computed using Fast Iterative Digital Image Correlation (FIDIC)\cite{barkochba2015} utilizing a 64 $\times$ 64 pixel subset, and a spacing of 16 pixels (10.4 {\textmu}m). Cell-substrate traction was computed through Fourier Transform Traction Cytometry with a correction for finite substrate thickness.\cite{trepat2009, butler2002, delalamo2007} Monolayer Stress Microscopy was used to calculate the stresses within the cell layer.\cite{tambe2011,tambe2013, saraswathibhatla2020sciData} This is done by applying the principle of force equilibrium to the cell-substrate traction, resulting in the stress tensor within the cell layer. The first and second principal stresses, $\sigma_1$ and $\sigma_2$, were computed by taking the maximal and minimal eigenvalues of the stress tensor, respectively. By convention, tensile (pulling) stress are positive, while compressive (pushing) stress are negative. 

Cell velocities, $\bm{v}$, were calculated using FIDIC on the phase contrast images of the cells, following the same subset size and spacing parameters described above. The strain rate tensor, $\dot{\bm\varepsilon}$, was computed according to $\dot{\bm\varepsilon} = [\nabla \bm{v} + (\nabla \bm{v})^T]/2$, where $\nabla \bm{v}$ represents the velocity gradient tensor, and $(\cdot)^T$ is the matrix transpose. The first and second principal strain rates, $\dot{\varepsilon}_1$ and $\dot{\varepsilon}_2$ correspond to the maximal and minimal eigenvalues of the strain rate tensor, respectively. 

\subsection{Calculation of effective viscosity}

To calculate the effective viscosity, the ratio of shear stress to shear strain rate was quantified at all locations in space using a moving window approach. A window size of 62$\times$62 {\textmu}m$^2$ was chosen such that good spatial resolution was achieved while keeping noise at a minimum, as described in our previous work.\cite{mccord2025arxiv} Within each window, 36 data points were used to compute the slope of the ratio of shear stress and shear strain rate via a linear mean square regression, and the slope was taken as the effective viscosity for that region. To improve spatial resolution, we overlapped adjacent windows by 10.4 {\textmu}m. The use of linear regression assumes a linear viscous behavior between stress and strain rate. Given that the rheology of the cytoskeleton is not linearly viscous,\cite{fabry2001, fabry2003} the assumption of linearity is not necessarily fully accurate but rather is an approximation. This assumption is reasonable, as the data are approximately linear over the range of strain rates studied here ($\approx 0.03 - 0.3$ hr$^{-1}$, Supplemental Fig. S4). We note that the assumed linearity should not be extrapolated to larger strain rates without independent verification. 

\subsection{Cell-cell adhesion experiments}
For cell-cell adhesion experiments, 1 mm cell islands were treated in 1$\times$ PBS for 30 min to facilitate antibody penetration at the cell-cell junctions. Then, cell culture medium containing 5 {\textmu}g/mL CD324 (E-Cadherin) Monoclonal Antibody (DECMA-1) (ThermoFisher 14-3249-82) suspended in control medium was added.\cite{choi2022} 
The 30 min treatment with PBS prevented performing the DECMA-1 experiment with the monolayer expansion experiment, because the monolayer expansion experiment required seeding the cells against the barrier for only 1 hr before beginning the experiment.

\subsection{Metabolic inhibition}
As described previously,\cite{mccord2025arxiv} metabolic inhibition experiments were performed in a solution of 135 mM NaCl, 5 mM KCl, 1.5 mM KCl2, 1 mM MGSO4, 10 mM HEPES, and 5.5 mM glucose. Metabolic inhibition was initiated by a spike of 10 mM 2-Deoxy-D-glucose (2-DG), and 2.5 mM NaCN to the solution.\cite{mccord2025arxiv, stelling2004,baron2005}

\subsection{Statistical testing}
Statistical analysis between groups was done using a Student's t-test. Statistical analysis on distributions was done using a Levene's test for equality of variances. For differences in slopes between groups, analysis of covariance was performed. The specific test used for comparison are noted in figure captions. Groups that had a value of $p=0.05$ or less were considered statically significant, with *,**, and *** indicating $p < 0.05$, $p < 0.01$, and $p < 0.001$, respectively.

\section*{Acknowledgments}
The authors thank Christian Franck for use of the confocal microscope. This work was supported by NSF grant CMMI-2205141 and NIH grant R35GM151171.

\section*{Code Availability}
Code used to analyze experimental data is available at https://github.com/jknotbohm/FIDIC and \\ https://github.com/jknotbohm/Cell-Traction-Stress.

\section*{Author Contributions}
J.N. and M.M. conceptualized the work. M.M. performed the experiments. J.N. and M.M. analyzed the experimental data and wrote the manuscript. 

\section*{Competing Interests}
The authors declare no competing interests.

\newpage
\bibliographystyle{unsrt-modified}
\bibliography{refs_positiveviscosity}

\newpage

\begin{center}

\textbf{Supplementary Information}

\Large

\textbf{Measurement of tissue viscosity to relate force and motion in collective cell migration}

\vspace{11pt}

\normalsize
Molly McCord$^{1,2}$, Jacob Notbohm$^{1,2,*}$
\vspace{11pt}

$^1$Department of Mechanical Engineering, University of Wisconsin--Madison, Madison, WI, USA\\
$^2$Biophysics Program, University of Wisconsin--Madison, Madison, WI, USA

$^*$ Contact: jknotbohm@wisc.edu

\end{center}

\vspace{18pt}
\section*{Supplemental Figures}
\vspace{11pt}

\setcounter{figure}{0} 

\renewcommand{\thefigure}{S\arabic{figure}} 

\begin{figure}[H]
\centering
\includegraphics[keepaspectratio=true, width=3.5in]{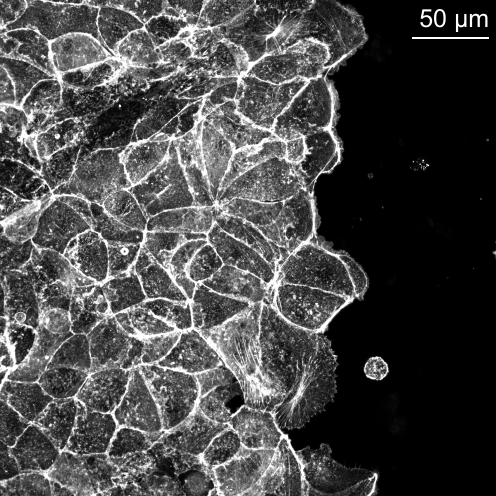}
\caption{Fluorescent labeling of the actin cytoskeleton at the edge of a cell monolayer at the time of barrier removal. The fluorescent intensity is similar at the edge of the monolayer and at cell-cell interfaces, indicating the absence of a multicellular actin cable.} 
\end{figure}

\begin{figure}[H]
\centering
\includegraphics[keepaspectratio=true, width=6in]{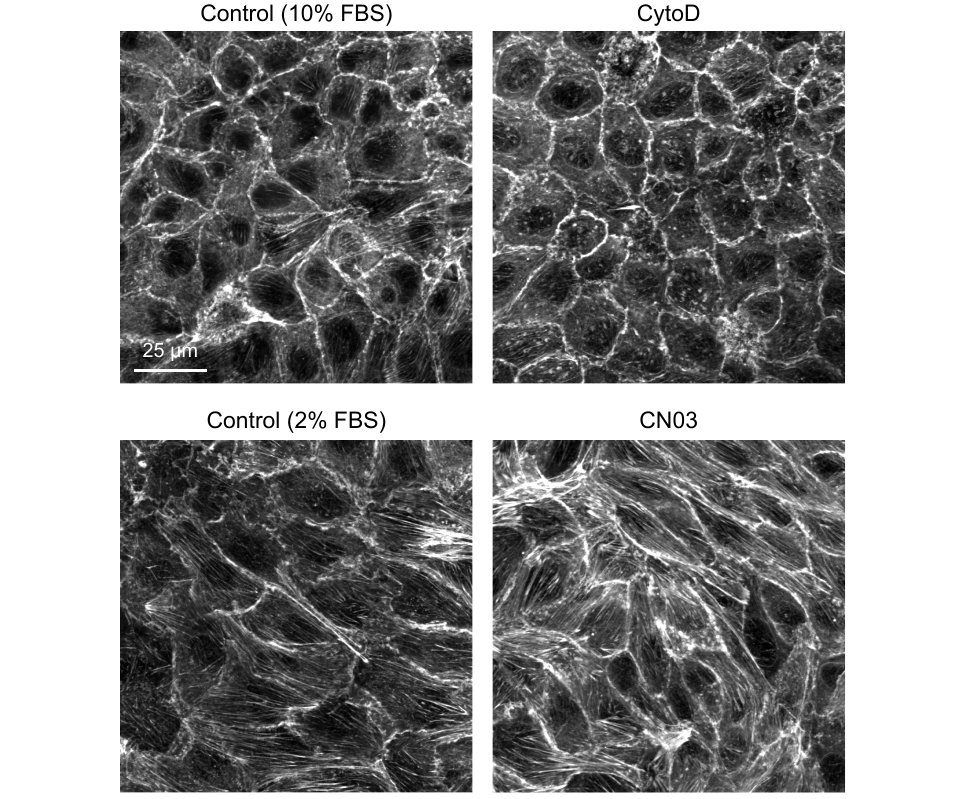}
\caption{Enlarged images of fluorescently labeled actin in response to cytochalasin D and CN03. Compared to their respective controls, treatment with cytochalasin D reduced the amount of stress fibers and treatment with CN03 increased the amount of stress fibers.}
\end{figure}

\begin{figure}[H]
\centering
\includegraphics[keepaspectratio=true, width=1.5in]{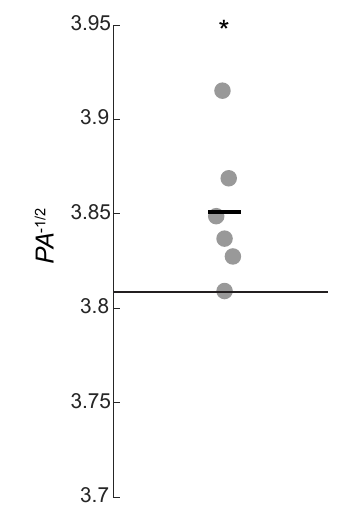}
\caption{Cell monolayers are within the viscous regime. Cells were segmented, and each cell's perimeter $P$ and area $A$ were measured. The quantity $P/\sqrt{A}$ was measured for each cell. Dots show averages of $P/\sqrt{A}$ over independent cell monolayers. Black bar indicates the mean. The data are statistically different from 3.81 ($p = 0.04$ one-sample t-test in comparison to 3.81), indicating the cells are in the fluid regime.
}
\end{figure}

\begin{figure}[H]
\centering
\includegraphics[keepaspectratio=true, width=6.5in]{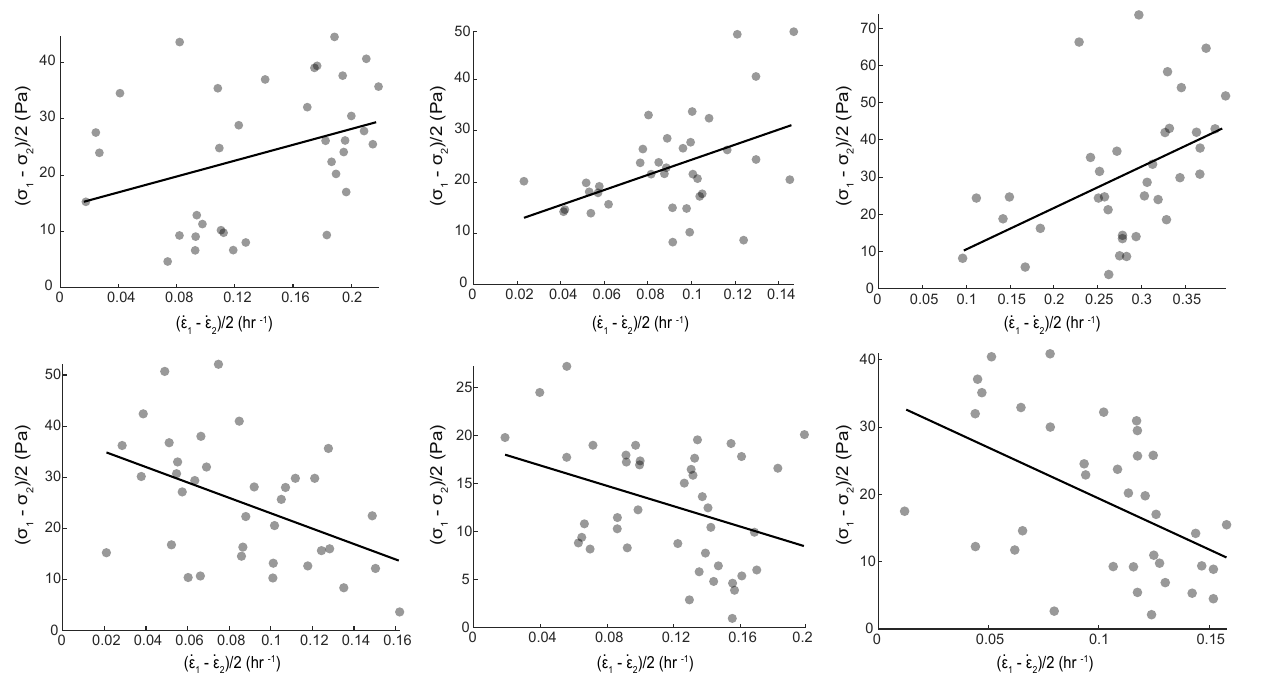}
\caption{Representative scatter plots of the relationship between shear stress and shear strain rate within different 62 $\times$ 62 {\textmu}m$^2$ windows chosen at random locations within one monolayer. Gray dots indicate points within each window, and the black lines indicate the linear fits.} 
\end{figure}

\begin{figure}[H]
\centering
\includegraphics[keepaspectratio=true, width=6.5in]{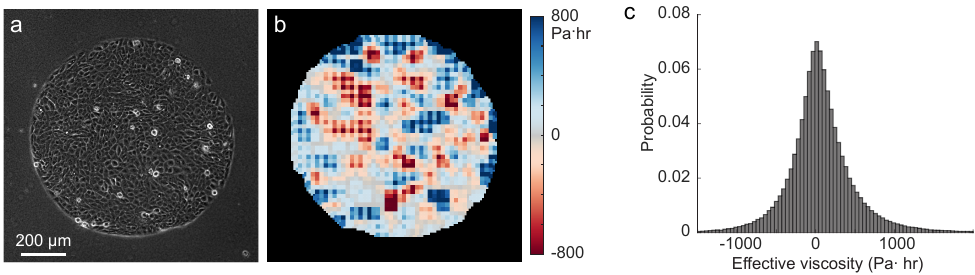}
\caption{Measurement of effective viscosity within HaCaT cell monolayers. (a) Representative phase contrast image of a HaCaT cell island. (b) Color map of effective viscosity. (c) Histogram of effective viscosity, showing a broad width and slightly positive mean. 
}
\end{figure}

\begin{figure}[H]
\centering
\includegraphics[keepaspectratio=true, width=6.5in]{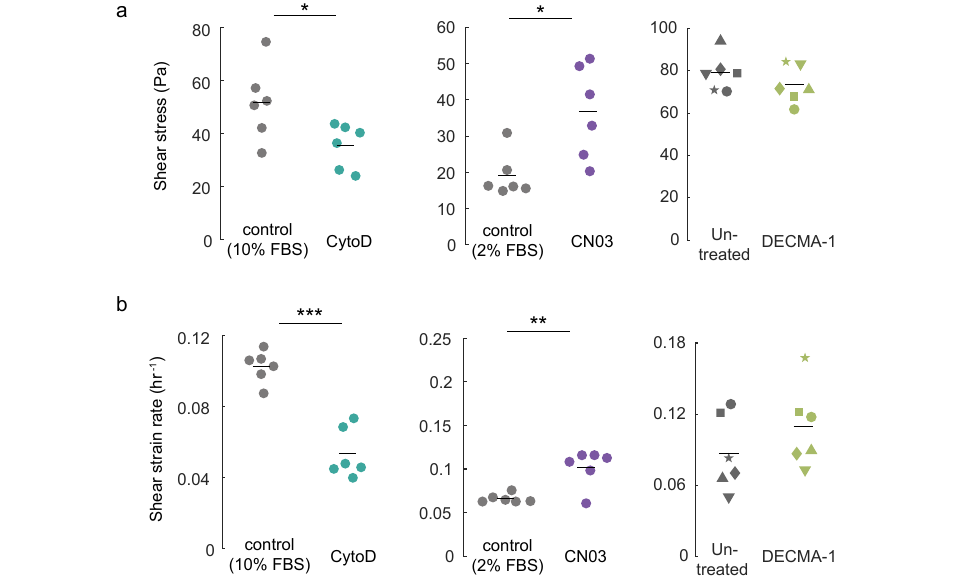}
\caption{Shear stress and shear strain rate for various treatments.
(a) Shear stress, $(\sigma_1-\sigma_2)/2$, for treatments and their respective controls (CytoD $p = 0.039$, CN03 $p = 0.012$, two-sample t-tests; DECMA-1 $p = 0.33$, one-sample t-test) Each dot represents a different cell island. For DECMA-1 treatment, the marker shape is the same cell island before and after treatment. 
(b) Shear strain rate, $(\dot\varepsilon_1-\dot\varepsilon_2)/2$, for treatments and their respective controls (CytoD $p < 0.0001$, CN03 $p = 0.002$, two-sample t-tests; DECMA-1 $p = 0.15$, one-sample t-test). Each dot represents a different cell island. For DECMA-1 treatment, the marker shape is the same cell island before and after treatment.}
\end{figure}

\begin{figure}[H]
\centering
\includegraphics[keepaspectratio=true, width=6in]{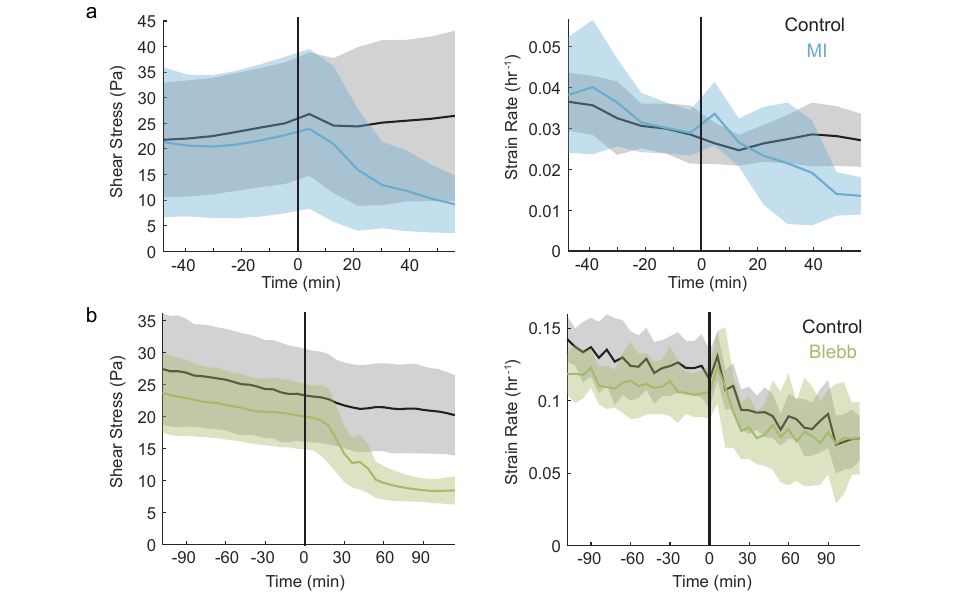}
\caption{Shear stress and shear strain rate before and after metabolic inhibition and treatment with blebbistatin. 
(a) Shear stress and shear strain rate over time in response to metabolic inhibition. Metabolic inhibition occurred at $t=0$. Solid lines indicate the mean and shading indicates the standard deviations over 6 (control) or 7 (treatment) cell islands. 
(b) Shear stress and shear strain rate over time in response to blebbistatin. Cells were treated with blebbistatin at $t=0$. Solid lines indicate the mean and shading indicates the standard deviations over 6 (control) or 7 (treatment) cell islands. 
Both treatments reduced the shear stress and strain rate, but both shear stress and strain rate remained above zero for the duration of the experiment.}
\end{figure}

\begin{figure}[H]
\centering
\includegraphics[keepaspectratio=true, width=6.5in]{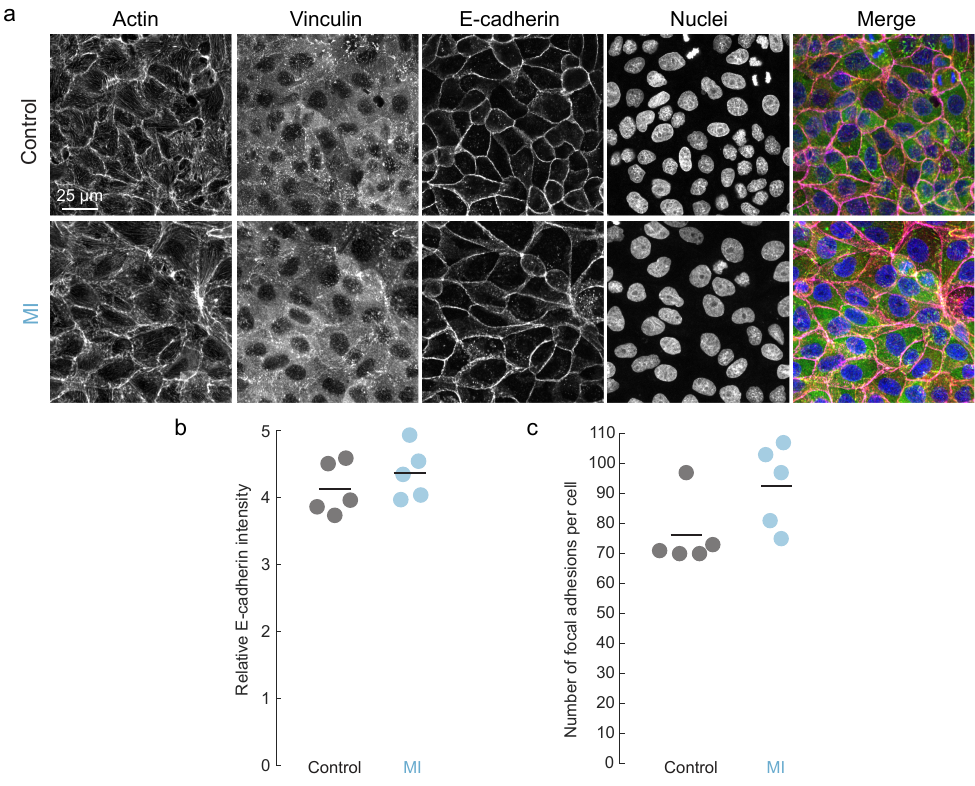}
\caption{Fluorescent labeling of the cytoskeleton, cell-substrate adhesions, and cell-cell adhesions for metabolic inhibition experiment. (a) Fluorescent images of actin, vinculin, E-cadherin, nuclei, and merged channels for control and metabolic inhibition groups. (b) Relative E-cadherin intensity and (c) average number of fofcal adhesions per cell for control and metabolic inhibition. Each dot represents the average over a field of view, and black bars indicate means.}
\end{figure}

\end{document}